\input{aipcheck}

\documentclass[
    ,final            
  ]
  {aipproc}

\layoutstyle{6x9}

\begin{document}

\def\apj {\emph{ApJ}}
\def\apjl {\emph{ApJ}}
\def\apjs {\emph{ApJS}}
\def\nat {\emph{Nature}}
\def\mnras {\emph{MNRAS}}
\def\aap {\emph{A\&A}}
\def\aapr {\emph{A\&A Rev.}}

\def\xmm {\emph{XMM-Newton}}
\def\cha {\emph{Chandra}}
\def\swi {\emph{Swift}}
\def\integral {\emph{INTEGRAL}}
\def\src {1E\,1547.0--5408}
\def\ee {1E\,1547.0--5408}
\def\flux {\mbox{erg cm$^{-2}$ s$^{-1}$}}
\def\lum {\mbox{erg s$^{-1}$}}
\def\nh {$N_{\rm H}$ }

\title{The spectacular X-ray echo of a magnetar burst}

\classification{97.60.Jd, 98.70.Qy, 98.38.Cp}
\keywords      {magnetars, X-ray bursts, interstellar dust, dust scattering}

\author{A.~Tiengo}{
  address={INAF/Istituto di Astrofisica Spaziale e Fisica Cosmica - Milano, via E.~Bassini 15, 20133 Milano}
}

\author{G.~Vianello}{
  address={INAF/Istituto di Astrofisica Spaziale e Fisica Cosmica - Milano, via E.~Bassini 15, 20133 Milano}
}

\author{P.~Esposito}{
  address={INAF/Istituto di Astrofisica Spaziale e Fisica Cosmica - Milano, via E.~Bassini 15, 20133 Milano}
,altaddress={INFN - Istituto Nazionale di Fisica Nucleare, Sezione di Pavia,
via A.~Bassi 6, 27100 Pavia, Italy}
}

\author{S.~Mereghetti}{
  address={INAF/Istituto di Astrofisica Spaziale e Fisica Cosmica - Milano, via E.~Bassini 15, 20133 Milano}
}

\begin{abstract}

The Anomalous X-ray Pulsar (AXP) \src\ reactivated in 2009 January with the emission of dozens of short bursts. Follow-up observations with \swi/XRT and \xmm\ showed the presence of multiple expanding rings around the position of the AXP. These rings are due to scattering, by different layers of interstellar dust, of a very high fluence burst emitted by \ee\ on 2009 January 22.
Thanks to the exceptional brightness of the X-ray rings, we could carry out a detailed study of their spatial and spectral time evolution until 2009 February 4. This analysis gives the possibility to estimate the distance of \ee. We also derived constraints on the properties of the dust and of the burst responsible for this rare phenomenon.

\end{abstract}

\maketitle

\section{Introduction}

\ee\ belongs to the small class of Anomalous X-ray Pulsars (AXPs),
which, together with the Soft Gamma-ray Repeaters (SGRs), are
thought to be magnetars, i.e. isolated neutron stars ultimately
powered by magnetic energy (see \cite{mereghetti08} for a
review). \ee\ is  the magnetar candidate with the shortest
spin period ($P=2.1$ s) and one of the two from which pulsations
have been detected also in the radio band
\cite{camilo07}.


On 2009 January 22, \ee\ was detected at its historically highest X-ray flux level and showed bursting activity of unprecedented frequency and intensity \cite{mereghetti09,savchenko09,kaneko09}.
Follow-up X-ray observations performed in the following 7 days with the
\swi\ satellite and on 2009 February 3--4 with \xmm\ led to the discovery of expanding rings
around \ee\ caused by interstellar dust scattering of an extremely intense burst of X-rays \cite{tiengo09gcn8848}.

The intensity of the dust scattering rings depends on the flux of
the emitted X-ray radiation and on the amount of dust in the line
of sight, while their angular size and flux time
evolution depend only on the distances of the source and of the dust
layers. Thus these distances can be derived even if the total optical
depth is unconstrained, provided that a dust scattering differential cross-section is known.
Here the main results of the analysis of the X-ray rings of \ee\ are reported; for a detailed description of the analysis and a deeper discussion, see \cite{tiengo09}.




\section{Data analysis, results and discussion}

During the observations of \ee, the X-Ray Telescope (XRT, \cite{burrows05}) on board \swi\ was operated
either in Windowed Timing (WT) mode
or in Photon Counting (PC) mode.
We  analyzed 5 observations in WT mode performed on 2009 January 22, with typical durations of $\sim$1--2 ks, and 7 observations in PC mode performed during the following week, with exposure times between $\sim$2 and $\sim$6 ks. The \xmm\ observation performed on 2009 February 3--4 was 50 ks long and all the EPIC detectors \cite{struder01,turner01} were operated in Full Frame (FF) mode
with the Thick optical blocking filter.

\begin{figure}
  \includegraphics[height=.3\textheight,angle=0]{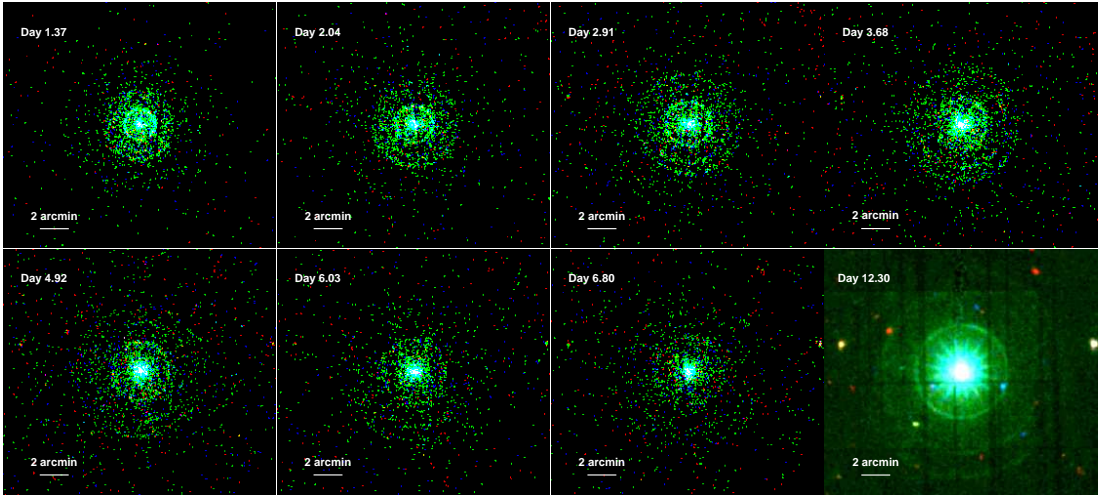}
  \caption{Images of the seven \swi/XRT observations in PC mode and of the \xmm/EPIC observation of \ee. The observation time (in days after the burst that likely produced the X-ray rings) is indicated in each panel.}
\end{figure}


Fig.~1 shows  the images of the XRT observations in
PC mode and of the EPIC observation, where three X-ray rings expanding with time are clearly visible.
To derive the expansion rate,
we measured the ring radii from each of these X-ray images by fitting the corresponding peaks in the \ee\ radial profiles.
Based only on geometrical considerations \cite{ts73}, the expansion rate of a ring formed by a short burst of X-rays scattered by a thin dust layer is given by:

\begin{equation}
\theta (t) = \sqrt{\frac{2 c}{d} \frac{(1-x)}{x} (t-t_0)} = K \sqrt{t-t_0} \label{thetat},
\end{equation}
where $\theta$ is the ring radius, $c$ is the speed of light,
$t_0$ is the burst time, $d$ is the source distance, and $x$ =
$d_{\mathrm{dust}}/d$ is the ratio of the dust layer and source distances.
The expansion coefficient $K$ is given by:
\begin{equation}
K = \sqrt{\frac{19.84 (1-x)}{x d}}, \label{kappa}
\end{equation}
where $\theta$ is in arcmin, $t$ in days, and $d$  in kpc.

We fit the expansion of the three rings with $K$ and $t_0$ as
free parameters, obtaining
$t_0$ values consistent with a single
epoch, while we obtained significantly different values for the expansion coefficients $K$ of the three rings. This indicates that the three rings are produced by the
scattering of the same burst by three different dust layers and
not by a single layer scattering three different bursts. In
particular the derived $t_0$ values are compatible with the time
of the strongest burst detected from \ee\ on 2009 January 22 by \integral\
\cite{mereghetti09,savchenko09} and \emph{RHESSI} \cite{bellm09}. This burst occurred at $T_0= 54853.28141$
MJD. Fitting jointly the three rings and imposing
a common $t_0$, we obtain a 90\% confidence interval $T_0-2100$ s $< t_0 < T_0 + 700$ s.
If we fix $t_0=T_0$, the best-fit expansion coefficients are:
$K_1=0.8845\pm0.0008$, $K_2=1.553\pm0.003$, and $K_3=2.000\pm0.002$ ($\chi^2$/dof=17.2/21; 1$\sigma$ errors).


The single-scattering halo profile (expressed in erg cm$^{-2}$ s$^{-1}$ keV$^{-1}$) predicted for a burst of fluence $F_{\rm X}$ (in erg cm$^{-2}$ keV$^{-1}$) scattered by  grains
with size distribution $n(a)$ (dust grains with radius $a$ per hydrogen atom)  is \citep{ml91}:
\begin{equation}
I(\theta,E)=3.6\times 10^{-5} \bigg(\frac{19.84+K^2d}{K~d}\bigg)^2 N_{\mathrm{H}} F_{\mathrm{X}} \int \mathrm{d} a ~ n(a)\frac{\mathrm{d}\sigma}{\mathrm{d}\Omega}, \label{ithetae}
\end{equation}
where $N_{\rm H}$ is the hydrogen column density in the dust cloud. In the Rayleigh-Gans approximation the differential scattering cross-section for the dust is:
\begin{equation}
\frac{\mathrm{d}\sigma}{\mathrm{d}\Omega}=1.1\bigg(\frac{\rho}{3}\bigg)^2 a^6 \Phi^2(\theta,E,a,K,d)~ \rm cm^2,  \label{sigma}
\end{equation}
where $\rho$ is the density (in g cm$^{-3}$) of each dust component and $\Phi(\theta,E,a,K,d)$ is the form factor. We used equation (\ref{ithetae}) to derive the distance of \ee\ and constrain the burst intensity.

The energy-dependent halo profiles, i.e.
the spectra $I(E,\theta)$ of the three rings at the different angles  $\theta$ from the
source direction, were extracted from all the available X-ray data by fitting the radial profiles (or the monodimensional images for the XRT data in WT mode) in different energy bands (see \cite{tiengo09} for details).
For the dust composition and grain size distribution, we considered many dust models: the one described in \cite{WD01}, the 15 models proposed in \cite{zda04} and an idealized model, where the dust is composed by a single dust component with a power-law grain size distribution $n(a)\propto a^{-\alpha}$, defined between $a_{\mathrm{min}}$ and $a_{\mathrm{max}}$ (see, e.g., \cite{mrn77}).

About half of the adopted dust models could not adequately fit the energy-resolved halo profiles of the three X-ray rings. The best-fit source distances obtained with the remaining models were in the 4--8 kpc range (see Table~2 in \cite{tiengo09}).
The best-fitting dust model (BARE-GR-B in \cite{zda04}) provides a distance of $3.91\pm0.07$ kpc for \ee, which is compatible with the proposed association with the supernova remnant (SNR) G\,327.24--0.13 \cite{gelfand07}. Moreover, applying equation (\ref{kappa}), this source distance implies distances of 2.2 kpc, 2.6 kpc and 3.4 kpc for the three dust clouds, in good agreement with the dust distribution inferred by CO line observations towards \src\ \cite{tiengo09}. However, considering the relatively large uncertainties on the distance estimates of the SNR and the dust clouds, a set of similarly well-fitting dust models that imply a source distance of $\sim$5 kpc cannot be excluded.
A distance of $\sim$4--5 kpc is also favored by the fact that all these dust models are already known to provide good fits to the dust-scattering halos of bright X-ray binaries (see, e.g., \cite{valencic09} and references therein).

The source distance does not depend on
the hydrogen column density in the dust clouds, which only affects the relative normalization of the prompt and scattered emission. Given that the total $N_{\rm H}$ measured in X-ray observations of \ee\ is $\sim$$3\times10^{22}$ cm$^{-2}$, to estimate the burst fluence we assume $N_{\rm H}=10^{22}$ cm$^{-2}$ in the dust cloud responsible for the brightest ring.
Assuming  a bremsstrahlung spectrum with $kT=100$ keV and any of the well-fitting dust models, we estimate that the burst producing the X-ray rings released an energy of
10$^{44-45}$ erg in the 1--100 keV band (see Table~3 in \cite{tiengo09}). Although our estimate is affected by large systematic uncertainties, it would mean that this burst was the brightest flare without any long-lasting pulsating tail ever detected from a magnetar.


Finally we note that during the day following the bright burst, the X-ray halo was brighter than the persistent emission of \ee.
This means that, if observed with an instrument with poor spatial resolution, the halo emission
would have been attributed to the persistent emission of \ee\ and
therefore interpreted as a burst X-ray afterglow.
Since all the Galactic magnetars are located at low latitudes, their X-ray emission is expected to pass through large amounts
of dust before reaching us; this means that some dust-scattering contribution
might significantly contaminate the
X-ray emission of magnetars following bright bursts.



\begin{theacknowledgments}
This research is based on observations with the NASA/UK/ASI \swi\
mission and with \xmm, an ESA science mission with instruments and
contributions directly funded by ESA Member States and NASA.
We acknowledge the partial support from ASI (ASI/INAF contracts
I/011/07/0, I/010/06/0, and I/088/06/0).
\end{theacknowledgments}



\bibliographystyle{aipproc}   

\end{document}